\documentclass[reprint,aip, jcp, intlimits, amsmath, amssymb]{revtex4-1}

\usepackage{graphicx}
\usepackage[caption=false]{subfig}
\usepackage[capitalize]{cleveref}
\usepackage{xcolor}
\usepackage{siunitx}
\usepackage{paralist}
\usepackage{multirow}

\begin{document}

\title{Modeling of effective interactions between ligand coated nanoparticles through symmetry functions}
\author{Dinesh Chintha}
\affiliation{Department of Chemical Engineering,Indian Institute of Science, Bangalore 560012, Karnataka, India }
\author{Shivanand Kumar Veesam}
\affiliation{Department of Chemical Engineering,Indian Institute of Science, Bangalore 560012, Karnataka, India }
\author{Emanuele Boattini}
\affiliation{Soft Condensed Matter, Debye Institute for Nanomaterials Science, Utrecht University, Princetonplein 5, 3584 CC Utrecht, The Netherlands}
\author{Laura Filion}
\affiliation{Soft Condensed Matter, Debye Institute for Nanomaterials Science, Utrecht University, Princetonplein 5, 3584 CC Utrecht, The Netherlands}
\author{Sudeep N Punnathanam}
\email{sudeep@iisc.ac.in}
\affiliation{Department of Chemical Engineering,Indian Institute of Science, Bangalore 560012, Karnataka, India }

\date{\today}

\begin{abstract}
Ligand coated nanoparticles are complex objects consisting of a metallic or semiconductor core with organic ligands grafted on their surface. These organic ligands provide stability to a nanoparticle suspension. In solutions, the effective interactions between such nanoparticles are mediated through a complex interplay of interactions between the nanoparticle cores, the surrounding ligands and the solvent molecules. While it is possible to compute these interactions using fully atomistic molecular simulations, such computations are too expensive for studying self-assembly of a large number of nanoparticles. The problem can be made tractable by removing the degrees of freedom associated with the ligand chains and solvent molecules and using the potentials of mean force (PMF) between nanoparticles. In general, the functional dependence of the PMFs on the inter-particle distance is unknown and can be quite complex. In this article, we present a method to model the two-body and three-body potentials of mean force between ligand coated nanoparticles through a linear combination of symmetry functions. The method is quite general and can be extended to model interactions between different types of macromolecules.
\end{abstract}

\maketitle

\section{Introduction}

Nanoparticles have unique size dependent electrical, optical, magnetic and chemical properties that differ from those found in bulk solids. Accordingly, materials formed by aggregates of nanoparticles are expected to possess novel and unique properties. This combined with the ability to tune the aggregation process, e.g. through compositional variation, leads to the promise of novel technologies in the fields of electronics, photonics,\citep{talapin2010} plasmonics, catalysis,\citep{kleijn2014, wang2010, kang2013a, kang2013b} sensing, etc. Self-assembly of nanoparticles represent one of the processes through which nanostructured materials can be synthesized. The mechanism of the self-assembly process is strongly determined by inter-particle interactions. Typical constituents of a nanoparticle include a metallic or semiconductor core surrounded by organic ligands grafted on to the core. This layer of ligands provide stability to a nanoparticle suspension through steric repulsion between nanoparticles. Suspensions of gold nanoparticles stabilized by thiol coated surfactants represent one of the most well studied systems of nanoparticles \cite{brust1994synthesis, sardar2009gold, kumari2019gold}. The surfactants are hydrocarbon chains in which one of the end consists of thiol group (SH) and another end is functionalized with proper functional groups which are chosen to tailor nanomaterials to get desired properties. The surfactants are chemically grafted to the gold nanoparticle via the Au-S chemical bond. 

Study of the self-assembly of nanoparticles requires simulations of large numbers of gold nanoparticles (of order $10^2 \text{--} 10^3$). Using fully atomistic models for describing the interactions in such simulations is still too expensive for present-day computers. To make the problem tractable, the typical strategy is to remove the many degrees of freedom involving the ligand chains and the solvent molecules and model the particles as a small number (typically one) of sites with an effective interaction that depends only on inter-particle distances. These effective interactions are computed from fully atomistic molecular simulations of nanoparticles. In this regard there have been a number of studies previously reported\citep{landman2004, schapotschinikow2009, bauer2017, liu2018, liu2019, liu2020, monego2018, travesset2017a, travesset2017b, zha2021, patel2007, schapotschinikow2008, kaushik2013, jabes2014, tang2017, baran2017, liepold2019, yadav2020, monego2020} that have computed effective inter-particle interactions. These studies looked at various aspects of interactions between nanoparticles including length of ligand chains, role of solvent, three-body interactions, etc. 

A seminal study on the interactions between ligand-capped nanoparticles in vacuum is the one by \citet{landman2004}. From their simulation results, they developed the Optimal Packing Model (OPM) that predicted the inter-particle spacing as a function of the ligand length and nanoparticle diameter. \citet{schapotschinikow2009} computed the three-body interactions between nanoparticles in vacuum. They found that for short ligand chains, the three-body interaction results in an energy penalty when capping layers overlap. However, for long ligand chains, the nanoparticles formed a chain which is energetically more favorable than close packing. They also developed another phenomenological model called the Overlapping Cone Model (OCM), which also accounted for many-body effects on the inter-particle spacing. \citet{bauer2017} found that three-body interactions in triplets of gold nanoparticles in vacuum are mainly repulsive and nearly independent of temperature. Atomistic simulations by \citet{liu2018} found that inter-particle spacing increased linearly with increasing ligand coverage. The same authors\citep{liu2019} also showed that the potential well depth scales linearly with increasing total length but it hardly depends on the particle size. The inter-particle distance depends strongly on the size of the nanoparticle but is weakly dependent on the total ligand length. \citet{liu2020} found that long ligand molecules round the shape of the nanocrystals and make the interaction nearly isotropic. In contrast, short ligand molecules leads to geometrically asymmetric morphology of the nanocrystals, so that the interaction is orientation-dependent. \citet{monego2018} discovered an unexpected inversion on the role of ligand length on colloidal stability of apolar nanoparticles. They found that increasing the ligand length increases colloidal stability in the core-dominated regime but decreases it in the ligand-dominated regime. Recently, \citeauthor{travesset2017a}\citep{travesset2017a, travesset2017b} developed the Orbifold Topological Model (OTM) that has improved upon both OPM and OCM. The predictions of the OTM were shown to be quite accurate for Binary Nanocrystal Superlattices (BNSLs) by \citet{zha2021}.

In experiments, nanoparticle self-assembly is typically carried out in suspensions. Consequently, there have been computational studies on the role played by the solvent towards effective nanoparticle interactions. \citet{patel2007} studied interaction between two thiol surfactant coated gold nanoparticles in supercritical ethane. They found that increasing the solvent density as well as making the ligands more branched caused increased repulsion between the nanoparticles. \citet{schapotschinikow2008} showed that a good solvent results in purely repulsive interactions. \citet{kaushik2013} found that hexane and toluene are ``good'' solvents for the nanoparticles and they penetrate the ligand corona all the way to the nanoparticle surface. \citet{jabes2014} showed that fluctuations within the ligand shell of thiolated gold nanoparticles give rise to a significant degree of anisotropy in effective pair interactions. Later studies by \citet{tang2017} and \citet{yadav2020} also highlighted the importance of anisotropy in the interactions between nanoparticles. The simulations of \citet{baran2017} showed that ligand mobility has only a small effect on the pairwise interactions between surfactant coated nanoparticles in a good solvent. The study on the effect of solvent on colloidal stability by \citet{monego2020} found that agglomeration is enthalpically driven and that, contrary to what one would expect from classical colloid theory, the temperature at which the particles agglomerate increases with increasing solvent chain length.

One of the challenges then, is to capture the two and three body interactions using a relatively simple fit that can then be exploited in large-scale coarse-grained simulations. These fits to two and three body interactions should satisfy the general features of interactions between surfactant coated nanoparticles in a good solvent. \citet{liepold2019} have given a detailed overview of these features of nanoparticle interactions and some of the important observations are given below.
\begin{enumerate}
\item The presence/absence of solvent plays a significant role in the interactions between two nanoparticles. The potential of mean force (PMF) between two dry nanparticles show a deep minimum indicating strong attraction.\citep{jabes2014} This attraction is due to interactions between the ligand chains. The depth of the potential minimum increases with increasing length of the ligand chains. The core-core interaction when the interparticle distance correspond to the minimum in the PMF is negligible compared to the overall interaction. 
\item When solvents are present, the nature of the two body interactions are significantly different. The interactions also depend on the nature of the solvent. If the organic ligands are soluble, then the solvent is termed as a good solvent, otherwise, it is termed as poor solvent. In presence of a good solvent, the ligand chains predominantly adopt an extended configuration whereas in a poor solvent, the ligand chains are in a collapsed and compact configuration. As a result, the two-body PMF between a pair of nanoparticles is always repulsive in presence of a good solvent, but has a deep minimum in presence of a poor solvent
\item There are significant three-body interactions between the nanoparticle due to fluctuations of the ligand chains. Previous estimates of three-body interactions between nanoparticle in vacuum show that they are mainly repulsive in nature.
\item In addition to three-body interactions, the fluctuations of the ligand chains also result in significant anisotropy in the instantaneous force between a pair of nanoparticles which can have a significant effect on self-assembly of nanoparticles.
\end{enumerate}

In recent years, a number of methods based on machine learning (ML) techniques have been proposed in order to approximate complex many-body interactions and predict the properties of molecules and materials based on a few reference calculations \cite{bartok2010gaussian,rupp2012fast,faber2016machine,behler2016perspective,glielmo2017accurate,grisafi2018symmetry}. Most of these techniques have been developed to speed up \emph{ab initio} molecular dynamics simulations, where the energy and forces are evaluated with very costly electronic structure methods. More recently, these techniques have also been employed to approximate the many-body interaction between colloidal particles decorated with a soft deformable shell \cite{boattini2020}.

In this paper, we compute the PMFs between a pair and triplets of gold nanoparticles coated with dodecanethiol surfactant chains in presence of supercritical ethane as the solvent. The system is similar to the one studied in \citet{jabes2014} with diameter of the gold core being 1.6 nm. Previous estimates of three-body PMFs between ligand coated nanoparticles have only been for systems in vacuum. Here we determine both the two-body and three-body interactions in solution. In order to be able to later use this interactions in further simulations, we fit the resulting interactions using a linear combination of the symmetry functions introduced by  \citeauthor{behler2007}\citep{behler2007,behler2011} 
There are many advantages in describing the inter-particle interactions using symmetry functions. Specifically, symmetry functions are designed to capture the local environment around a particle in a way that takes into account the symmetry of the particle and its interactions; they are invariant to translations and rotations of the particle coordinates, and invariant to the exchange of particles of the same species. Additionally, they are analytic, continuous and differentiable which enables one to easily incorporate them into simulations where either energies or forces are required. 
As shown in Ref. \citenum{boattini2020}, it is possible to model complex many-body interactions among colloidal particles through linear regression of the symmetry functions. The same procedure is used in this work to model the two-body and three-body inter-particle interactions between surfactant coated gold nanoparticles.

\section{Simulation Details}
\subsection*{Forcefield}
The forcefield for simulating the system of dodecanethiol covered 1.6 nm gold nanoparticles immersed in supercritical ethane is same as the one used in Ref. \citenum{jabes2014}. The dodecanethiol surfactant chain and the ethane solvent are modeled using the united atom representation with CH$_2$, CH$_3$ and SH groups as the pseudo-atoms. The non-bonded interaction between the pseudo-atoms is modeled by the Lennard-Jones potential, i.e.,
\begin{align}
U_\text{LJ} (r)& = 4 \epsilon \left[ \left(\frac{\sigma}{r}\right)^{12} - \left(\frac{\sigma}{r}\right)^{6} \right] \notag
\end{align}
The Lennard-Jones parameters values for non-bonded interaction between unlike groups are obtained using the Lorentz-Berthelot rules. The parameter values are taken from the SKS forcefield and given in \cref{tab:LJ_SKS}
\begin{table}[h!]
\centering
 \begin{tabular}{c @{\extracolsep{1cm}} c  c} 
 \hline\hline
  pseudo-atom & $\epsilon$ (kJ/mol) &  $\sigma$ (\AA) \\ 
 \hline
 CH$_3$  & 0.9478  & 3.93  \\ 
 CH$_2$  & 0.3908 & 3.93  \\
 SH   & 1.6629  & 4.45 \\
 \hline \hline
 \end{tabular}
 \caption{Lennard-Jones force field parameters for ligand-ligand, ligand-solvent and solvent-solvent interactions}
 \label{tab:LJ_SKS}
\end{table}
The pseudo-atoms are connected by a rigid bond of length $l_0$, the bond angles are modeled by the harmonic potential and the torsional interactions are given by the triple cosine potential, i.e.,
\begin{widetext}
\begin{align}
U_\text{angle} (\theta) & = \frac{k_\theta}{2} \left( \theta - \theta_0\right)^2 \notag \\
U_\text{torsion} (\phi) & = \frac{a_1}{2} \left[1 + \cos(\phi)\right] + \frac{a_2}{2} \left[ 1 - \cos(2\phi)\right] + \frac{a_3}{2} \left[ 1 + \cos(2\phi)\right] \notag 
\end{align}
The values of the parameters in the interaction are given in \cref{tab:bonded}. 
\end{widetext}

\begin{table}
\begin{tabular}{l|l@{\extracolsep{5 mm}}ccc}
\hline\hline
\multirow{3}{*}{Bond} & pseudo-atoms& $l_0$ \\ \cline{2-5}
         & CH$_x$-CH$_x$ & 1.54 \\
         & CH$_2$-SH & 1.82 \\ \hline \hline
\multirow{3}{*}{Angle} &pseudo-atoms & $k_\theta$  & $\theta_0$ \\ \cline{2-5}
    & CH$_2$-CH$_2$-CH$_x$ & 519.73 & 114 \\
    & CH$_2$-CH$_2$-SH          & 519.73 & 114 \\ \hline \hline
\multirow{3}{*}{Torsion} & pseudo-atoms& $a_1$ & $a_2$ & $a_3$  \\ \cline{2-5}
    & CH$_2$-CH$_2$-CH$_2$-CH$_x$ & 5.9046 & -1.1340 & 13.1608 \\ 
    & CH$_2$-CH$_2$-CH$_2$-SH          & 5.9046 & -1.1340 & 13.1608 \\ \hline
\end{tabular}
\caption{Values of the parameters for various bonded interactions of the dodecanethiol ligand chain. The units of length, angle and energy are in \AA, degrees and kJ/mol respectively.}
\label{tab:bonded}
\end{table}

Previous studies\citep{landman2004, schapotschinikow2009} have shown that the PMF between a pair of nanoparticles in vacuum shows a deep minimum. At this distance, which can be considered as the closest distance of approach between a pair of nanoparticles, the interaction between the gold cores is negligible. Hence, following \citenum{jabes2014}, we omit direct interaction between the gold atoms in the simulations. The interaction of the gold atoms with the ligand and solvent atoms is modeled using a spherically averaged representation of the gold core. Again, following Ref. \citenum{jabes2014}, the entire gold core is modeled as a single site and its interaction with the other pseudo-atoms is via a modified $m-n$ functional form, i.e.,
\begin{align}
U(r)=4\epsilon\left[\left(\frac{\sigma}{r-r_0}\right)^{m} - \left(\frac{\sigma}{r-r_0}\right)^n\right] \notag
\end{align}
where parameters are given in \cref{tab:forcefield2}. The non-bonded interactions are all truncated and shifted at a cut-off distance of 15 \AA. All the simulations discussed in this paper was performed using the molecular dynamics package GROMACS.\citep{abraham2015} The density of the ethane solvent was 0.4525 g/cc. All the calculations were performed at a temperature of 300 K.

\begin{table}
\centering
 \begin{tabular}{l @{\extracolsep{5 mm}} c c c c c} 
 \hline \hline
 pseudo-atoms & m & n & $\epsilon$  & $\sigma$  & $r_0$  \\ \hline
 \hline\hline
 Au-CH$_3$ & 12  & 4  & 2.5033 & 3.051  & 5.32 \\ 
 Au-CH$_2$ & 12 & 4  & 1.6075 & 3.051 & 5.32 \\
 Au-SH & 12  & 10  &  581.076  & 3.43 & 5.50 \\
 \hline
 \end{tabular}
 \caption{Force field parameters for interaction between gold core and pseudo-atoms of ligands and solvent. The unit of $\epsilon$ is kJ/mol and that of $\sigma$ and $r_0$ is \AA.}
 \label{tab:forcefield2}
\end{table}

\begin{figure}
\centering
\includegraphics[width=3in]{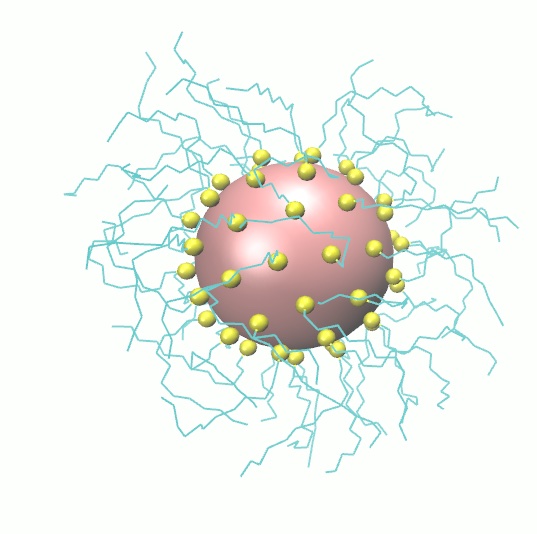}
\caption{Visualization of a thiol coated gold nanoparticle used in this study. The pink sphere represents the gold core, the yellow spheres represent the sulfur atoms and the cyan lines represent the hydrocarbon chains.}
\label{fig:single_np}
\end{figure}

\subsection*{Passivation of Gold Nanoparticle}
The starting configuration of dodecanethiol coated gold nanoparticle was created in the following sequence of steps. In step one, a gold cluster was kept at the center of a cubic simulation box of length 80 \AA. In the second step, 300 dodecanethiol molecules were inserted in the box around the gold nanoparticle. All the positions of dodecanethiol molecules are generated by using Packmol package.\citep{martinez2009} While generating the positions of these dodecanethiol molecules, we ensure that they do not overlap with each other. In the third step, the system was equilibrated by performing a molecular dynamics simulation at 300 K for a period of 5 ns. At the end of this equilibration run, 58 dodecanthiols were attached to the surface of the gold nanoparticle. In the fourth and final step, the values of $\epsilon$ given in \cref{tab:forcefield2} was increased by 20\% and an additional molecular dynamics simulation was performed for another 5 ns. This resulted in a total of 62 dodecanethiol chains attached to the gold core. A snapshot of the thiol coated nanoparticle generated from this procedure is shown in \cref{fig:single_np}. These passivated nanoparticles are used in simulations to compute the two-body and three-body interactions in both vacuum and solvent. The procedure to immerse the nanoparticles in the solvent is described in the results section.

\section{Modelling}
Since it is computationally expensive to determine the two-body and three-body interactions between nanoparticles using fully atomistic molecular simulations, it would be useful to have simple and easy to compute analytical functions that can describe these interactions. These functions describing the effective interactions can then be applied to study the behavior and properties of a large collection of nanoparticles such as during their self-assembly. To model the two-body interaction between any two nanoparticles, $u^{(2)}$, separated by a distance $r$, we use the symmetry function, $G^{(2)}$, which provides information on the pair correlation between two particles. Here we follow Ref. \citenum{boattini2020} and model the inter-particle interaction using a linear combination of these symmetry functions.
Accordingly we express $u^{(2)}$ as
\begin{align}
u^{(2)} (r) &= \sum_{i=1}^{N_s} \alpha_i G^{(2)}_i (r) \\
\label{eq:u2}
G^{(2)}_i (r) & = e^{-\eta_i (r-r_{0,i})^2}f_c(r)
\end{align}
where $N_s$ is the total number of symmetry functions used, and the parameters $\eta$ and $r_0$ control the width and the position of the Gaussian respectively. The function $f_c$  is a cutoff function that decreases monotonically and smoothly goes to 0 in both value and slope at the cutoff distance $r_c$. Note that a cutoff function is included to account for the finite range of the model.  Additionally, the cutoff function guarantees that the fitted function and its derivative smoothly go to zero at a specific distance, which is not only a desirable physical property, but also convenient for later use in simulations. 
The form of this cutoff function is given by
\begin{equation}
f_c(r) = 
\begin{cases}
0.5 \left[\cos\left(\pi \frac{r}{r_c} \right)+1\right] & \text{for}\ r\leqslant r_c \\
0 & \text{for}\ r > r_c
\end{cases}
\end{equation}

The modeling of the three-body interactions, $u^{(3)}$, for a triplet of nanoparticles requires symmetry functions that provide information on angular correlations in addition to distance correlations. This is done through the following functions:
\begin{eqnarray}
G^{(3)}(\textbf{r}_i, \textbf{r}_j, \textbf{r}_k) &=& 2^{1-\xi}(1+\lambda \cos \theta_{ijk})^\xi \times  \\ && \nonumber e^{-\eta\left(r^2_{ij}+r^2_{jk}+r^2_{ki}\right)}f_c(r_{ij})f_c(r_{jk})f_c(r_{ki}) \\
G^{(4)}(\textbf{r}_i, \textbf{r}_j, \textbf{r}_k) &=& 2^{1-\xi}(1+\lambda \cos \theta_{ijk})^\xi \times \\ &&  \nonumber e^{-\eta\left(r^2_{ij}+r^2_{jk}\right)}f_c(r_{ij})f_c(r_{jk})
\end{eqnarray}
where the angle $\theta_{ijk}$ is the angle formed by vectors $\textbf{r}_{ji}$ and $\textbf{r}_{jk}$, and $\xi$, $\eta$ and $\lambda$ are the parameters of the symmetry functions. To make the symmetry functions invariant to all permutations among the triplet of particles, we then define the following functions.
\begin{eqnarray}
Y^{(2)} (\textbf{r}_1, \textbf{r}_2,\textbf{r}_3) &=& G^{(2)}(r_{12}) G^{(2)}(r_{23})G^{(2)}(r_{31}) \\
Y^{(3)} (\textbf{r}_1, \textbf{r}_2,\textbf{r}_3) &=& G^{(3)}(\textbf{r}_1, \textbf{r}_2, \textbf{r}_3) G^{(3)}(\textbf{r}_2, \textbf{r}_3, \textbf{r}_1) \times  \nonumber\\ && G^{(3)}(\textbf{r}_3, \textbf{r}_1, \textbf{r}_2)  \\
Y^{(4)} (\textbf{r}_1, \textbf{r}_2,\textbf{r}_3) &=& G^{(4)}(\textbf{r}_1, \textbf{r}_2, \textbf{r}_3) G^{(4)}(\textbf{r}_2, \textbf{r}_3, \textbf{r}_1)  \times  \nonumber \\ && G^{(4)}(\textbf{r}_3, \textbf{r}_1, \textbf{r}_2)
\end{eqnarray}
The three-body interactions are now modeled as a linear combination of the symmetry functions $Y^{(j)}$ as follows 
\begin{equation}
u^{(3)} (\textbf{r}_1, \textbf{r}_2, \textbf{r}_3) = \sum_{i=1}^{N_s} \alpha_i Y^{(j)}_i(\textbf{r}_1, \textbf{r}_2, \textbf{r}_3)
\label{eq:u3}
\end{equation}
where $j\in \{2,3,4\}$. Note that $u^{(3)}$ corresponds to a family of functions with different values of $j$, $\xi$, $\eta$ and $\lambda$.

\section{Results and discussion}
\subsection{Two-body interaction}
\begin{figure}
\centering
\includegraphics[width=\linewidth]{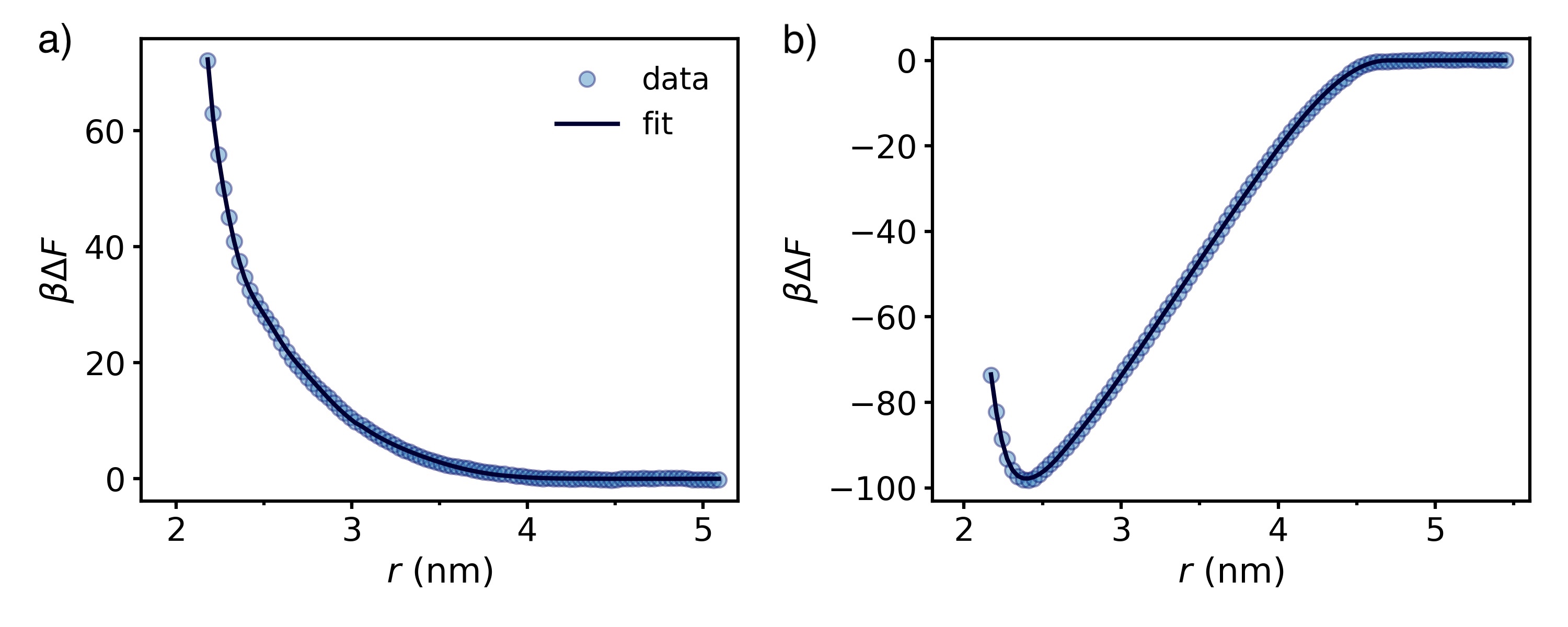}
\caption{Potential of mean force between two nanoparticles when (a) immersed in ethane solvent and (b) in vacuum. The symbols are data obtained from the simulations and the line is the prediction from the model based on symmetry functions.}
\label{fig:2body}
\end{figure}

To begin, we compute the effective two-body interaction, i.e., the PMF between a pair of ligand coated nanoparticles immersed in ethane, using atomistic molecular dynamics simulations. The starting configuration for these simulations was prepared as follows. A cubic simulation box of length 12 nm containing 31392 molecules of ethane, corresponding to a density of 0.4525 g/cc was prepared initially. Into this simulation box, two ligand coated nanoparticles having configurations generated by the method described above were introduced. One of the particles was kept at the origin and the other is kept a certain distance away along the x-axis. After insertion of the nanoparticles, all the ethane molecules that overlapped with the nanoparticles were removed from the box. During simulations, the center of particle one was kept fixed at the origin and the center of particle two was allowed to move along the x-axis only. In order to improve sampling, a series of simulations were performed with an external biasing potential to control the distance between the two particles. The biasing potential has the form $U_\text{bias} (r) = \frac{k_\text{b}}{2} \left( r - r_0\right)^2$, where $r$ is the distance between the two nanoparticles. The values of $k_\text{b}$ and $r_0$ were chosen in such a way so as to sample distances between 2 and 5 nm. Each molecular dynamics simulation consisted of an equilibration stage of 3 ns and a production stage of 17 ns. The values of $k_\text{b}$ varied between 1000-3500 kJ/mol/nm$^2$. The potential of mean force (PMF) between the two nanoparticles was computed using the weighted histogram analysis method (WHAM). The PMF computed from simulations is shown in \cref{fig:2body} (a). In agreement with the results reported in \citet{jabes2014}, the potential is repulsive throughout. This is as expected since ethane is a good solvent. We also computed the PMF between the two nanoparticles in vacuum and the computed values are shown in \cref{fig:2body} (b). In vacuum, there is a deep minimum in the PMF at around $r=2.4$ nm. This shows a strong attraction between the nanoparticles in vacuum which acts as poor solvent. This result is in agreement with \citet{jabes2014}. The equilibrium separation distance can also be predicted from phenomenological models such as OPM\citep{landman2004} and OCM\citep{schapotschinikow2009}. When these models are applied to our system, the OPM and the OCM predicted an equilibrium separation of 3.04 nm and 1.82 nm respectively. In these calculations, we have taken the value of the ligand length to be 1.56 nm from \citet{landman2004}. In an interesting coincidence, the value of 2.4 nm seen in our simulation is nearly equal to the mean of the values predicted by OPM and OCM. 

\subsubsection{Fitting with symmetry functions}
Having calculated the effective pair interaction between nanoparticles, we now model this interaction using a linear combination of the symmetry functions described in the previous section. 
Accordingly, the PMF values obtained from simulations are fitted to the following equation
\begin{equation}
    w(r) = \sum_{i=1}^{N_s} \alpha_i G_i^{(2)}(r) + c
    \label{eq:2body_fit}
\end{equation}
where $w(r)$ is the value of the PMF between two particles separated by a distance $r$ and $c$ is an arbitrary constant. The presence of $c$ in the above equation is due to the fact that only the relative difference between the values of PMF at two different distances have a physical meaning. During regression, we determine the values of both $\alpha_i$'s and $c$, but discard the value of $c$ in the model for $u^{(2)}$ as in \cref{eq:u2}. A similar strategy is also used to determine the coefficients in \cref{eq:u3}. A value of $r_c=4.7$nm was chosen as the cutoff radius.

The most important step in the fitting procedure is the selection of a suitable set of symmetry functions for use in \cref{eq:2body_fit}. To achieve this selection, we follow the procedure described in \citet{boattini2020}, who modified a similar procedure described earlier by \citet{imbalzano2018}. In this procedure, one starts with a large pool of symmetry functions. As described by \citet{imbalzano2018}, this pool spans all meaningful sets of parameters, using simple heuristic rules to represent most of the possible correlations within the cutoff distance. Guidelines for choosing this parameter set are given in \citet{imbalzano2018}. In the next step, subset of these functions that captures the most relevant features of the particle’s environment are selected from the pool one after the other in a way that maximizes the overall correlation with the target interparticle interaction. A brief explanation of the process for selecting the best set of symmetry functions is as follows. Let $u_i$ be the PMF for the $i^{th}$ data and $S_k(i)$ be the corresponding prediction by the $k^{th}$ symmetry function. The Pearson correlation coefficient\citep{imbalzano2018, boattini2020} is defined as
\begin{equation*}
c_k = \frac{\sum_i \left(S_k(i)-\bar{S_k}\right)\left(u_i-\bar{u}\right)}{\sigma_{SD}(S_k)\sigma_{SD}(u)}
\end{equation*}
where $\bar{S_k}$ and $\bar{u}$ are the arithmetic mean of $S_k$ and $u$ respectively, and 
$\sigma_{SD}(S_k)$ and $\sigma_{SD}(u)$ are the corresponding standard deviations. The first symmetry function that is chosen is the one with the highest value of $c_k$. The remaining symmetry functions are chosen one by one in the following manner. The linear correlation between a set of SFs and the energy is quantified by the coefficient of multiple correlation, $R$, whose square is given by
\begin{equation*}
R^2 = \mathbf{c}^\text{T}\mathbf{R}^{-1}\mathbf{c}
\end{equation*}
where, $\mathbf{c}^\text{T} = (c1 , c2 , \ldots)$ is the vector whose $k^{th}$ component is the Pearson correlation, $c_k$, while $\mathbf{R}$ is the correlation matrix of the current set of symmetry functions. Specifically, the element $R_{km}$ of this matrix is the Pearson correlation between $S_k$ and $S_m$ in the set. More simply, $R^2$ is a measure of the square correlation between predictions and actual energies, and has a maximum value of 1 for perfect predictions. Note, $R^2$ does not depend on the chosen units and can be used to assess the general quality of the fit.

\begin{figure}
\centering
\includegraphics[width=\linewidth]{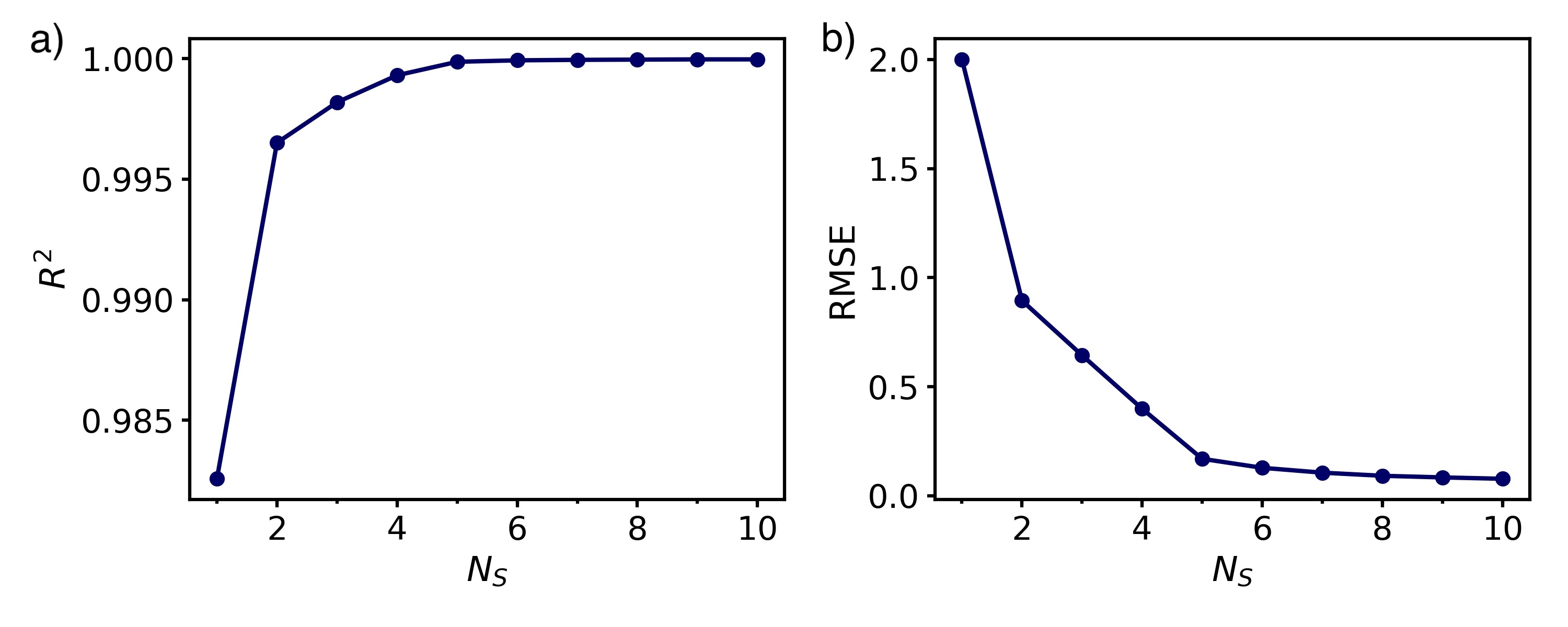}
\caption{(a) Square of the correlation coefficient, $R^2$, and (b) root mean squared error (RMSE) as a function of the number of symmetry functions employed, $N_s$ for fitting the two-body interactions between nanoparticles in ethane solvent.}
\label{fig:2body_Rsq}
\end{figure}

For two-body interaction, we start from a pool of 470 symmetry functions where $\eta \in [2^{-2},2^7]$ in geometric progression with a factor of 2 and $r_0 \in [0,4.7]$ in arithmetic progression with a factor of 0.1. \Cref{fig:2body_Rsq} shows plots of the coefficient of multiple correlation, $R^2$, and the root mean squared error of the linear fit (RMSE) for the two-body interaction. Note that RMSE value depends both on the quality of the fit and the chosen units. $R^2$ and the RMSE are related by a simple transformation, i.e., $R^2 = 1 - \text{RMSE}^2 /\sigma_{SD}^2 (u)$. We observe that the value of $R^2$ increases substantially between $N_s=1$ and $N_s=2$, followed by small but noticeable increases till $N_s=5$. Any further increase in the value of $R^2$ is quite small. The decrease in the value of RMSE mirrors this increase in the value of $R^2$. Based on these results, the value of $N_s$ was set equal to 5, since any further increase in $N_s$ will only result in a small improvement in the fit. The values of $\eta$, $r_0$ and $\alpha$ for this choice of $N_s$ is given in \cref{tab:u2} and the fit is shown in \cref{fig:2body} (a). We also perform a similar regression for PMF in vacuum and the predictions from the model are shown in \cref{fig:2body} (b). The fitted model is found to be in excellent agreement with the simulation data.

\begin{table*}
\begin{tabular}{c|@{\extracolsep{1cm}}rrr|rrr}
\hline\hline
&\multicolumn{3}{c|}{Ethane} & \multicolumn{3}{c}{Vacuum} \\
\hline
$i$ & $\alpha$ & $\eta$ & $r_0$ & $\alpha$ & $\eta$ & $r_0$ \\
\hline
1 & 337.43 & 0.5 & 0.8 & -367.38 & 0.25 & 3.8\\
2 & -27.79 & 16  & 2.4 & -3.13   & 31.25 & 2.5 \\
3 &  10.43 & 4   & 3.4 & -197.54 & 1.25 & 5.0 \\
4 & 1372.71 & 64  & 1.9 & 112.81 & 31.25 & 2.0 \\
5 &  6.68 & 64  & 2.5 & 66.60 &  1.25 & 3.3 \\
\hline
\end{tabular}
\caption{Values of the coefficients and the parameters [\cref{eq:u2}] of the symmetry functions used to model the two-body interactions between nanoparticles.}
\label{tab:u2}
\end{table*}

\subsection{Three-body interaction}

\begin{figure}
\centering
\includegraphics[scale=0.4]{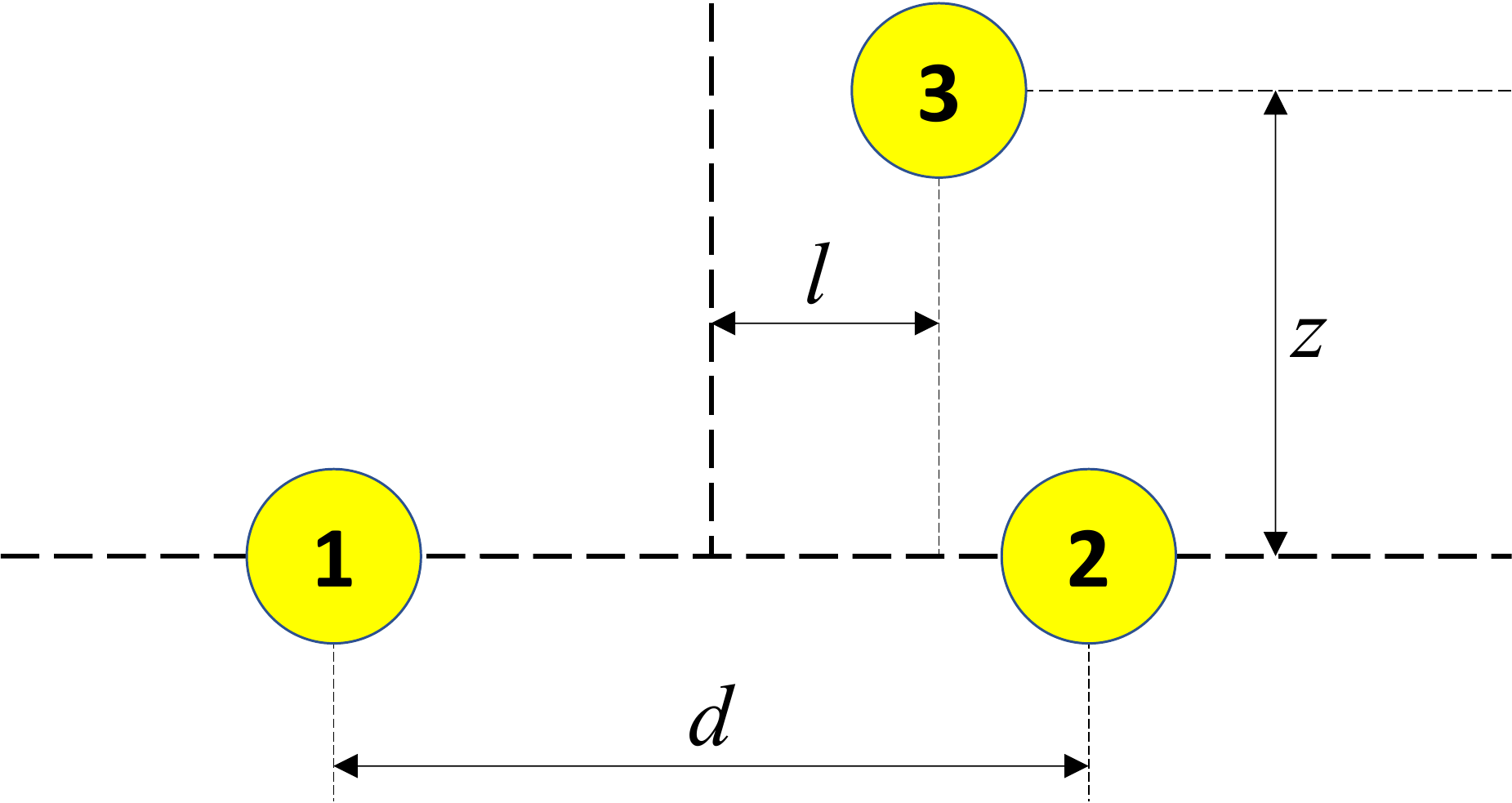}
\caption{Schematic showing the relative placement of a triplet of nanoparticles for computing the effective three-body interaction. Particles labelled 1 and 2 are kept fixed while particle labelled 3 is moved perpendicular to the line connecting particles 1 and 2.}
\label{fig:3body_schematic}
\end{figure}

\begin{figure}
\centering
\includegraphics[width=\linewidth]{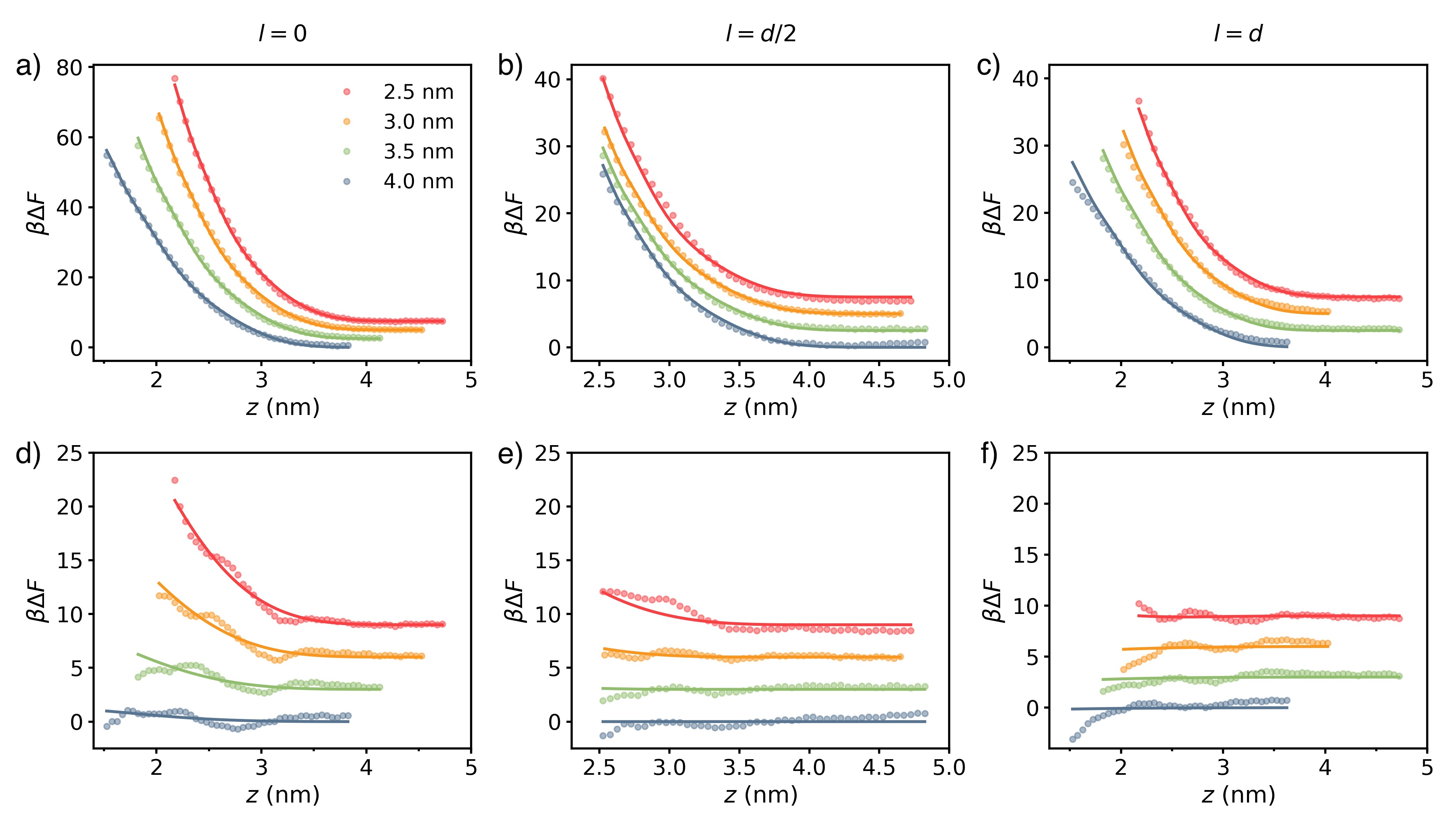}
\caption{Potentials of mean force felt by particle 3 shown in \cref{fig:3body_schematic} from the other two particles in ethane solvent for value of $d$ equal to 2.5 nm, 3.0 nm, 3.5 nm and 4.0 nm. The top row shows the total potential of mean force while the bottom row shows the interaction solely due to three body effects. Note that the points indicate simulation data while the lines are the fits. The different curves have been shifted vertically for clarity.}
\label{fig:3body_eth}
\end{figure}

Similar to the computation of two-body PMFs, the effective interactions between triplets of nanoparticles were computed from atomistic molecular dynamics simulations. In order to sample the various distances and angles between a triplet of nanoparticles, we use the following strategy. A schematic showing the relative positions of the three nanoparticles are shown in \cref{fig:3body_schematic}. During these simulations, two nanoparticles are kept fixed with a distance, $d$, separating their centers. The third particle is moved in a direction that is perpendicular to the line connecting the first two particles. This particle is at a distance of $l$ from the plane bisecting the line-segment connecting the first two particles. A series of simulations were performed with an external biasing potential of the form $U_\text{bias} (z) = \frac{k_\text{b}}{2} \left( z - z_0\right)^2$, where $z$ is the distance of particle 3 from the line connecting particles 1 and 2. Each individual simulation consisted of three nanoparticles and 31392 molecules of ethane. The length of the cubic simulation box was 12 nm. The starting configuration for these simulations is prepared similar to the one for computing two-body interactions. Each molecular dynamics simulation consisted of an equilibration stage of 8 ns and a production stage of 32 ns. The PMF computed from these simulations gives the total interaction between a triplet of nanoparticles. This data is shown in the top row of \cref{fig:3body_eth} for three different values of $l$. The three-body interactions between the nanoparticles is then estimated by subtracting the two-body contributions using \cref{eq:u2} and are shown in the bottom row of \cref{fig:3body_eth}. These data show that while the relative strength of the three-body interactions are much smaller than the two-body interactions, they do nevertheless make a significant contribution to the overall interaction and should be included while studying systems of nanoparticles. The three-body interactions are mostly repulsive except for certain configurations which will be discussed later. The strength of the three-body interaction is strongest when the third particle lies in the plane bisecting the line connecting the first two particles (see \cref{fig:3body_eth} (d)).

\begin{figure}
\centering
\includegraphics[width=\linewidth]{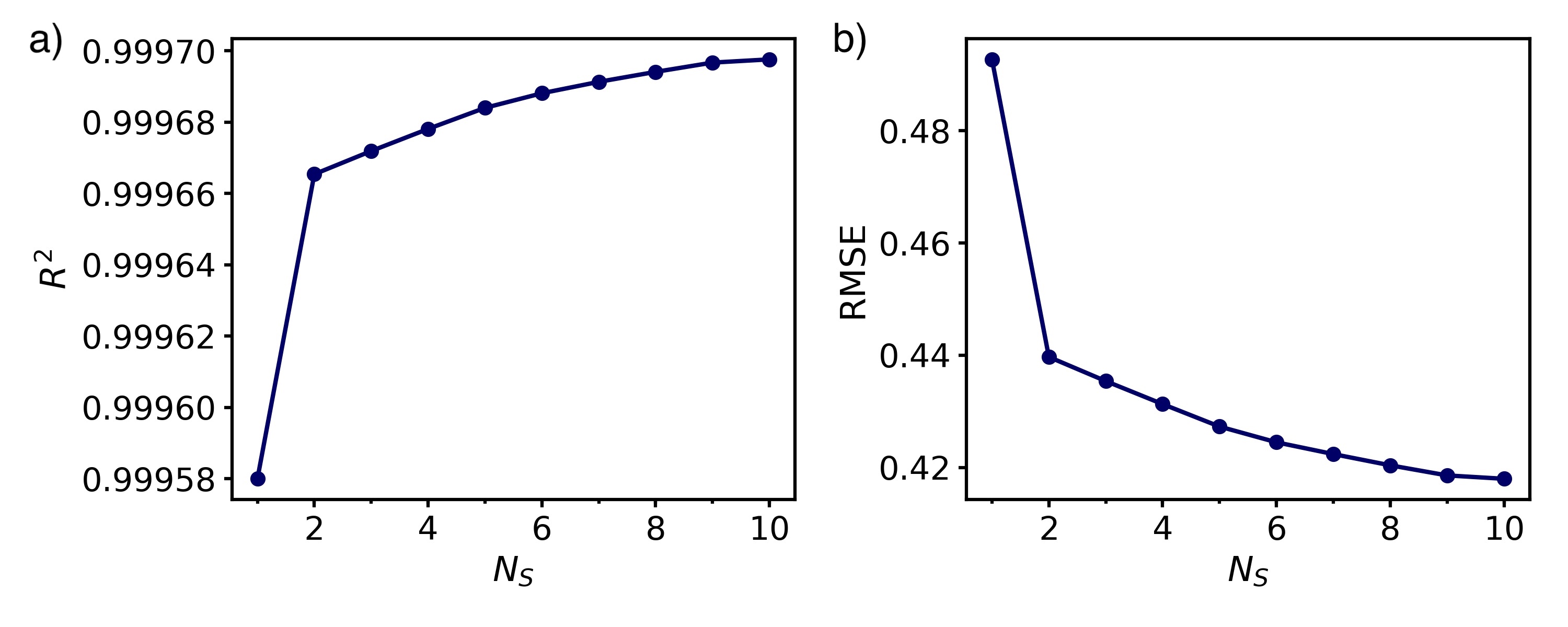}
\caption{(a) Square of the correlation coefficient, $R^2$, and (b) root mean squared error (RMSE) as a function of the number of symmetry functions employed, $N_s$ for fitting the three-body interactions between nanoparticles in ethane solvent.}
\label{fig:3body_Rsq}
\end{figure}

\begin{table}
\begin{tabular}{c|@{\extracolsep{0.5cm}}c|rrccc}
\hline\hline
$i$ & $j$ & $\alpha$ & $\eta$ & $r_0$ & $\xi$ & $\lambda$\\
\hline
1 & 2 & 149.56 & 0.5 & 1.5 & - & - \\
2 & 4 & -1.52  & -0.1 & - & 8 & -1 \\
\hline
\end{tabular}
\caption{Values of the coefficients and the parameters [\cref{eq:u3}] of the symmetry functions used to model the three-body interactions between nanoparticles.}
\label{tab:u3}
\end{table}

\subsubsection{Fitting with symmetry functions}

\begin{figure}
\centering
\includegraphics[width=\linewidth]{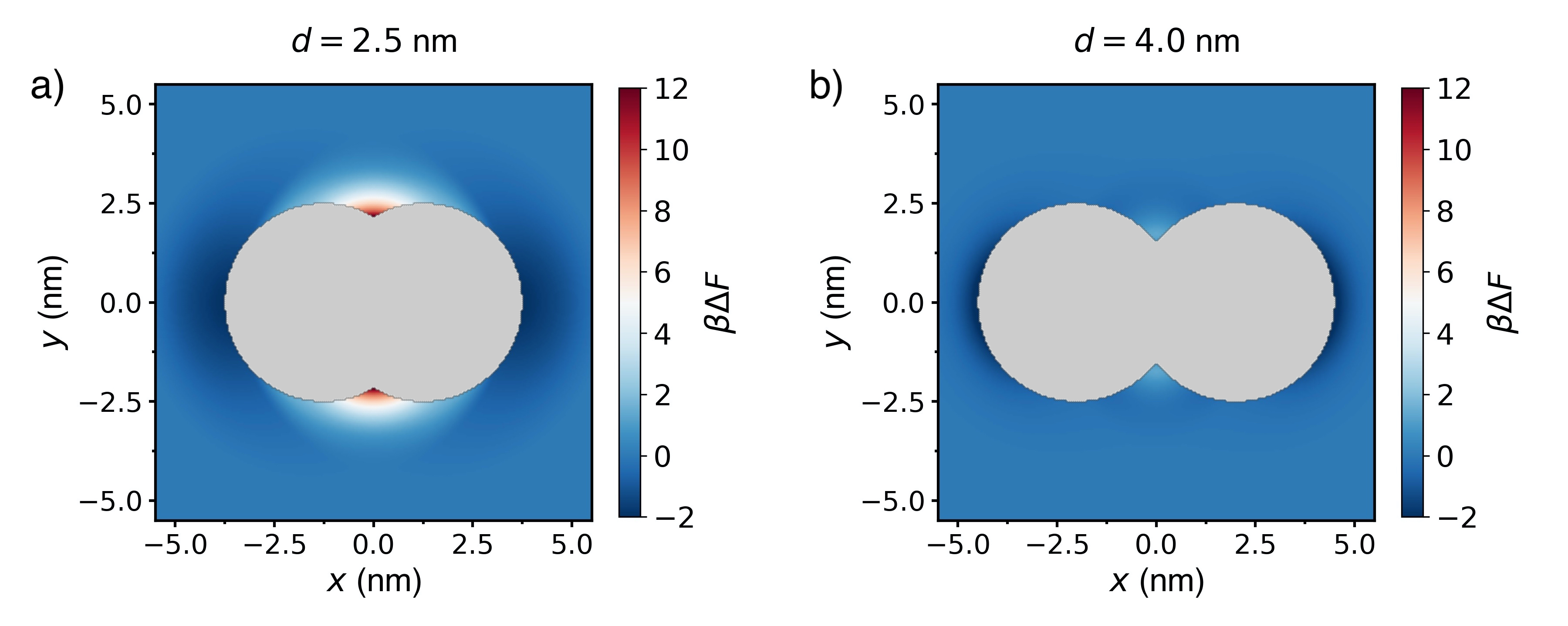}
\caption{Color map depicting the three body interactions, $u^{(3)}$, between a triplet of surfactant coated nanoparticles as a function of the location of the third particle. The first two nanoparticles are located at the center of the two overlapping grey circles which are separated by a distance $d$. The radius of a grey circle is 2.5 nm which is the distance at which the two body potential is approximately $25\ k_BT$. So the probability of the third particle coming any closer is negligible.}
\label{fig:int_map}
\end{figure}

For modeling the three-body interactions, a pool of 447 symmetry functions were considered for linear regression. For $G^{(2)}$ function, we chose $\eta \in [0.05,6.4]$ in geometric progression with a factor of 2 and $r_0 \in [0, 4.7]$ in arithmetic progression with a difference of 0.1. For $G^{(3)}$ and $G^{(4)}$ functions, we chose  $\eta \in -[0.0001,1]$ in geometric progression with a factor of 10, $\xi \in [1,32]$ in geometric progression with a factor of 2 and $\lambda \in \{-1,1\}$. The corresponding plots of the coefficient of multiple correlation, $R^2$, and the root mean squared error of the linear fit (RMSE) are shown in \cref{fig:3body_Rsq}. Although the values of $R^2$ are close to 1, the increase in the value of $R^2$ between $N_s=1$ and $N_s=2$ is substantial compared to further increases. Compared to the two-body interactions, however, the fit for the three-body interactions show a higher value of RMSE. This is due to higher error in the computed value of the three-body interactions. We set the number of symmetry functions used for fitting the three body interactions to $2$. Any further increase in the number of symmetry functions did not result in any noticeable improvement in the fit.  The values of the parameters are shown in \cref{tab:u3} and the predictions from this model is compared with the simulation data in \cref{fig:3body_eth}. \Cref{fig:int_map} shows  color maps of the entire three-body interactions for two different values of $d$. 

\begin{figure}
\centering
\includegraphics[width=0.6\linewidth]{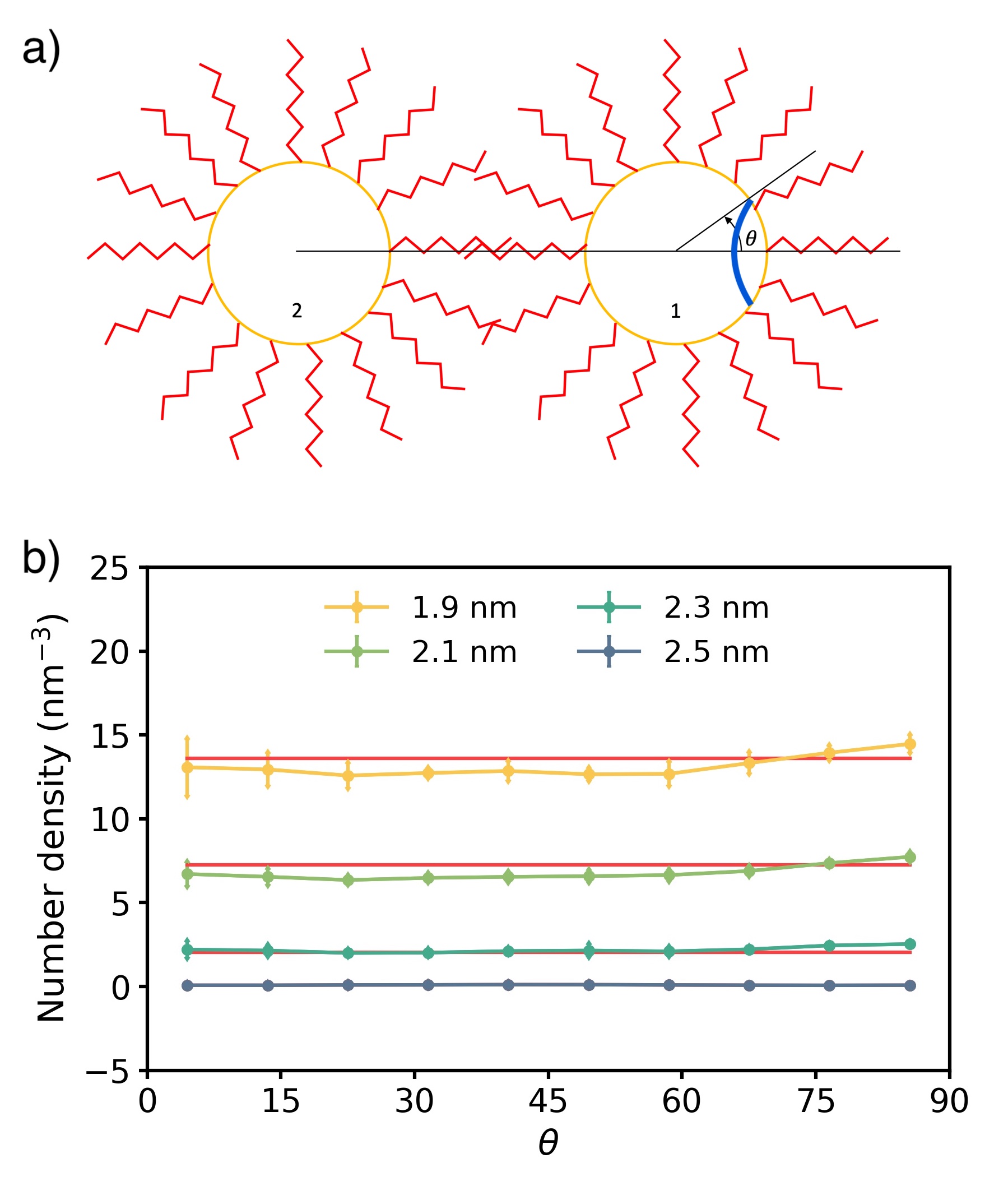}
\caption{(a) Schematic of the two nanoparticles and the definition of the angle $\theta$. The number density of the surfactant atoms is computed inside a circular shell (shown in the blue color). (b) Number density of surfactant atoms surrounding a nanoparticle at various distances from its center inside the circular shell. The red lines show the data for an isolated nanoparticle. The other lines show number density around nanoparticle in the presence of a second nanoparticle.}
\label{fig:density_profile}
\end{figure}

As previously stated, the computed three-body interactions are mostly repulsive. This would imply that the repulsion between two particles would increase due to the presence of a nearby third particle. This observation of three-body repulsion is in agreement with finding of \citet{schapotschinikow2009}. However, if we observe the \cref{fig:int_map} closely, we find that the three body interaction becomes negative, i.e, attractive, along the axis connecting the centers of the two nanoparticles. This is most prominently seen when $d=2.5$ nm. The magnitude of the three-body interaction is still much smaller than the two-body interaction and hence the overall interaction is still repulsive, but the repulsion is reduced due to the presence of the third particle. This is a surprising result since we had expected the three-body interaction to be positive throughout all configurations. To understand the origin of this negative $u^{(3)}$, we computed the number density of the atoms of the surfactant chains surrounding a nanoparticle with and without the presence of a third nanoparticle. The computed number density is shown in \cref{fig:density_profile}. The number density is computed for a circular shell (see \cref{fig:density_profile} (a)). The line joining the centers of the two nanoparticles passes through the center of this circular shell and is perpendicular to the plane of this circular shell. Every point on this circular shell forms an angle $\theta$ with the line joining the centers of the two nanoparticles. For an isolated nanoparticle, the number density at fixed distance from the center of the particle, as expected, is uniform. When a third particle is brought near, then there is a slight reduction in the number density in the range $\theta \in (0,80)$. We attribute this lowering of the number density to the attraction between the surfactant chains of the two nanoparticles. The reduction in the number density of the surfactant chains will result in a reduced repulsion between the nanoparticles. This is reflected in the negative values for the three-body interaction. This reduction in the interaction is quite significant (around 2 $k_B T$). It would be interesting to study the effect of this three-body attraction on the self-assembly of nanoparticles.

\section{Conclusion}
In this work, we have performed molecular simulations of thiol surfactant coated gold nanoparticles immersed in supercritical ethane that acts as a good solvent. From these simulations, we have computed the two-body and three-body interactions among the nanoparticles. These are expensive and time consuming calculations and any study involving large number would require an easy to compute fast model for these interactions. To this end we we fit the simulation data using a linear combination of symmetry functions. 
This linear regression was able to represent fairly accurately the qualitative and quantitative features of the nanoparticle interactions. The success of the symmetry function approach to model complex interactions between surfactant coated nanoparticles opens up the possibility to study phenomena involving large numbers of such nanoparticles such as their phase behavior and self-assembly.

\section*{Data Availability Statement}
The data that support the findings of this study are available from the corresponding author upon reasonable request.

\acknowledgements
The funding for this work has been provided by the Department of Science and Technology, Government of India (Grant No. DST/INT/NL/P-02/2016/C) and the Netherlands Organisation for Scientific Research (NWO) (Grant No. 16DDS004). The computations were carried out using computers purchased under the Nano Mission Programme of Department of Science and Technology, Government of India (Grant No. DST/NM/NS-14/2011(C)).

\bibliography{Manuscript}

\end{document}